\documentclass[11pt]{article}

\usepackage{jheppub,tabu}
\usepackage{package-notes}
\usepackage{NewCommands}
\usepackage{comment}
\usepackage[utf8]{inputenc} 

\def\gca{\mathfrak{gca}}
\def\bms{\mathfrak{bms}}

\title{The complex null string, Galilean conformal algebra and scattering equations}
\author[1,4]{Eduardo Casali,}
\author[2,5]{Yannick Herfray}
\author[3,4]{\&~Piotr Tourkine}

\begin{document}

\preprint{DAMTP-2017-32\\}

\affiliation[1]{The Mathematical Institute,\\ University of Oxford, Woodstock Road, Oxford OX2 6GG, UK}
\affiliation[2]{Laboratoire de Physique, \\ ENS Lyon, 46 allée d’Italie, F-69364 LYON CEDEX 07, FRANCE}
\affiliation[3]{Department of Applied Mathematics and Theoretical Physics,\\ University of Cambridge, Wilberforce Road,
  Cambridge CB3 0WA, UK}
\affiliation[4]{Kavli Institute for Theoretical Physics\\
University of California, Santa Barbara, CA 93106, USA}
\affiliation[5]{School of Mathematical Sciences,\\  University of Nottingham,
University Park, Nottingham NG7 2RD, UK}

\emailAdd{casali@maths.ox.ac.uk}
\emailAdd{yannick.herfray@ens-lyon.fr}
\emailAdd{pt373@cam.ac.uk}

\abstract{The scattering equation formalism for scattering amplitudes,
  and its stringy incarnation, the ambitwistor string, remains a
  mysterious construction.
  In this paper, we pursue the study a gauged-unfixed version of the
  ambitwistor string known as the null string.
  We explore the following three aspects in detail; its
  complexification, gauge fixing, and amplitudes.  We first study the
  complexification of the string; the associated symmetries and
  moduli, and connection to the ambitwistor string. We then look in
  more details at the leftover symmetry algebra of the string, called
  Galilean conformal algebra; we study its local and global action and
  gauge-fixing. We finish by presenting an operator formalism, that
  we use to compute tree-level scattering amplitudes based on the
  scattering equations and a one-loop partition function.
  These results hopefully will open the way to understand conceptual
  questions related to the loop expansion in these twistor-like string
  models.}

\maketitle

\section{Introduction}
\label{sec:introduction}

One of the most recent striking developments in the study of
scattering amplitudes is the discovery of the Cachazo-He-Yuan (CHY)
formalism \cite{Cachazo:2013iea,Cachazo:2013hca} for massless
scattering in field theory. The CHY formalism recasts scattering
amplitudes in terms of contour integrals in the complex
plane based on the solutions to the \textit{scattering equations}.

These contour integrals, reminiscent of the twistor string
\cite{Witten:2003nn}, were shown to originate from a new class
of string theories dubbed 'ambitwistor strings'~\cite{Mason:2013sva}.
These allowed the extension of the original CHY formulae in many
directions; loops \cite{Adamo:2013tsa,Ohmori:2015sha,Adamo:2015hoa}, curved
backgrounds \cite{Adamo:2014wea,Azevedo:2016zod,Chandia:2015sfa,Chandia:2015xfa}, manifestly supersymmetric versions \cite{Berkovits:2013xba,Chandia:2016dwr}, and even a string field theory
\cite{Reid-Edwards:2015stz,Reid-Edwards:2017goq}.

However, some basic aspects of this formalism remain unexplained, such
as its gauged-unfixed form and the connection to standard string
theory.
Particularly at loop-level questions related to modular invariance
and the integration domain are still not settled
\cite{Casali:2014hfa}. The extension of some recent developments at
one and higher loops \cite{Geyer:2015bja,Geyer:2015jch,Geyer:2016wjx}
may rely on a deeper understanding of these questions.

In \cite{Casali:2016atr}, two of us argued that the ambitwistor
string's origin is a theory partially characterized in the literature
called null strings. This theory was initially introduced by Schild
\cite{Schild:1976vq} as the \emph{classical} tensionless limit of the
usual string theory sigma-model. 

The idea that ambitwistor strings, describing only massless field
theory scattering, could be related to a tensionless limit of string
theory is actually counter-intuitive, some evidence for it was present in \cite{Bandos:2014lja,Bandos:2017zap} but was not developed further. In \cite{Casali:2016atr} it was
emphasized that this is only a classical statement. Quantum
mechanically, it is a remarkable quantization ambiguity, already
discovered in the 80's~\cite{Gamboa:1989zc,Gamboa:1989px}, that
truncates the spectrum of the string to a finite number of states,
essentially the massless sector of the usual string (see also
\cite{Hohm:2013jaa,Huang:2016bdd,Leite:2016fno,Hohm:2016lim}).\footnote{Another
  choice of quantization yields a theory more compatible with what is
  expected from the high energy limit \cite{Gross:1987ar}.}

The goal of this paper is to build up on the work done in
\cite{Casali:2016atr} in three directions, making more
precise the relationship of this theory to the the CHY formalism. In particular we
hope that this should open the way to a deeper understanding of the
loop expansion of these models.
The main results we provide are:
\begin{itemize}
\item A study of the complexification of the null string, its
  symmetries and moduli. These we match with the ambitwistor
  string. Understanding the global structure of this moduli space will
  eventually lead to a proper determination of the integration cycle
  of the ambitwistor string at loop-level, along the lines of
  \cite{Witten:2012bh,Ohmori:2015sha}. 
    
\item We use the representation theory of the constraint algebra of
  the string, called Galilean Conformal (GCA)\cite{Bagchi:2009pe,Bagchi:2013bga,Bagchi:2015nca,Bagchi:2016yyf}, to show how the
  chirality of the string emerges due to decoupling of null states.
  We characterize its action on the moduli and the match the zero
  modes determinant with the ghost determinant from ambitwistor
  string.  {This gives a new perspective on the truncation of the
    spectrum and its chirality. }
\item We propose a new computation of tree-level amplitude and
  one-loop partition function using operator methods. The scattering
  equations emerge thanks to the integration of the original 'time'
  coordinate of the string, an idea originally due to
  \cite{Siegel:2015axg}. We conjecture on modular transformations.
\end{itemize}

These three results are discussed in sections
\ref{sec:complexification}, \ref{sec:gca} and \ref{sec:op},
respectively. The sections are mostly self-contained and can be read
independently. 

\section{The complex null string}\label{sec:complexification}

\subsection{From the null string to the ambitwistor string}

The null string was originally obtained by Schild as a tensionless
limit of the Nambu-Goto string \cite{Schild:1976vq}. The equivalent
second order form of this action on which this work is based is the  Lindstr\"om-Sundborg-Theodoridis (LST) action
\cite{Lindstrom:1990qb,Lindstrom:1990ar,Karlhede:1986wb,Isberg:1993av}:
\begin{equation}
  \label{eq:LST}
  S=\int d^2\sigma  V^\alpha V^\beta \partial_\alpha X^\mu \partial_\beta X^\nu G_{\mu\nu}
\end{equation}
where $G$ is the target space-time metric that we take to be flat
$G_{\mu\nu}=\eta_{\mu\nu}$, $X^\mu(\sigma,\tau)$ are the coordinates
of the string, and $V^\alpha$, $\alpha=\{0,1\}$ is a vector field with
density weight $(-1/2,-1/2)$.

Th light-cone gauge and BRST quantization of the null string was done
in the seminal work \cite{Gamboa:1989zc}. To the best of our
knowledge, it was observed there for the first time that a
quantization ambiguity linked to the ordering of the operators leads
to two very different quantum theories: a higher-spin type one, still
poorly understood, and the one of interest for us, which is essentially the same as the ambitwistor string.

In this quantization the spectrum is truncated to the massless modes of string theory, and
although the bosonic model has negative-norm states, the
supersymmetric version is well-defined and its spectrum is the same as type II
supergravity. For a more complete review of the null string,
see~\cite{Casali:2016atr}, where the relation of the null string to
the ambitwistor string was studied.

In this section, we come back on a geometrical aspect that was not
discussed in this reference linked to the complexification of the
model. Indeed, the LST action is a real one, while the ambitwistor
string is a complex model.
In a longer term perspective, understanding the complexified model, in
loops for instance, will crucially rely on understanding the
complexification from the real model itself\cite{Witten:2010cx,Witten:2010zr}.

So let us describe step-by-step what we call the complex null string, its
geometrical meaning and symmetries.

We first allow the target space to be a complex manifold
$M^D_{\C}$ of (complex) dimension $D$, as well as allow the worldsheet
field $V$ to take complex values. That is, $V$ takes values in the
complexified tangent space to the worldsheet. At this point
the worldsheet itself is still a two dimensional real manifold.
This procedure gives a complexified version of the LST action where
$X : \gS \mapsto M^D_{\C} $ and
$V \in \left(\mathrm{\gO^2(\gS)}\right)^{\frac{1}{2}} \otimes T_{\C}
\gS$ are respectively, a map from the worldsheet to complexified
Minkowski space $M^D_{\C}\simeq \C^D$, and a complex vector field on
the worldsheet with weight one half.

Because $V$ is complexified, it generically defines a complex structure by requiring $V \in T^{(0,1)}\gS$, i.e.\footnote{This $\pab$ operator should be interpreted as a worldsheet field, depending on moduli, and not as a fixed background structure.}
\begin{equation}
V \propto \pab_{\zb}.\label{eq:V-dbar}
\end{equation}
Equivalently, it defines a conformal structure on $\gS$ through
  \begin{equation}
\tilde{g}^{\ga \gb}:= V^{(\ga} \overline{V}^{\gb)} 
\label{eq:conf-str}
\end{equation}
where $\overline{V}$ is the complex conjugate of $V$. In the real case, i.e. $V=\overline{V}$, this metric is degenerate as is usual in the null string.

We can therefore think of a choice of $V$ as a choice of complex structure together with a choice of ``scaling". We now discuss the interpretation of this ``scaling'' part. Let,
\begin{equation}
 V = \left( \frac{dz d\zb}{e}\right)^{\frac{1}{2}} \otimes \pab_{\zb}.
\label{eq:e-transformation}
 \end{equation} 
Keeping $V$ fixed while making a holomorphic change of coordinates $z \mapsto f\left(z \right)$ gives the following transformation law for ``$e$'':
\begin{equation}
e \mapsto e (\pa_z f)(\pab_{\zb} \bar{f})^{-1}.
\end{equation}
This implies that for a given $V$ we can think of the field $e$, as
the coordinates of a Beltrami differential: 
\begin{equation}
e\; d\zb \otimes \pa_{z}.\label{eq:e-Beltrami}
\end{equation}
As a consequence we have the following geometrical interpretation: If $\M$ is the space of complex structures on $\gS$ then a choice of $V$ is equivalent to choosing a point in $\Gamma := T\M$.

A quick look at the LST action, now written in terms of complex structure and Beltrami differential,
\begin{equation}\label{Second Order ambitwistor string}
	S\left[\pab, e, X \right] = \int_{\gS} \frac{dz d\zb}{e}(\pab X)^2,
\end{equation}
is enough to see that this is exactly the second order version of the ambitwistor action described in \cite{Mason:2013sva}:
\begin{equation}
S[\pab,e,X,P] = \int dzd\zb \left(P\cdot\pab X - \frac{e}{2} P\cdot P
\right).\label{eq:dbar-field}
\end{equation}
Note that in this action, the complex structure is a field of the
model, being integrated over, while the ambitwistor string is already gauge-fixed to conformal gauge.

\subsection{Equations of motion and boundary term}

To obtain the equations of motion we vary the action with respect to $X$\footnote{Here and everywhere below $\sqrt{d\gs^2}\;(VX)^{\mu}$ stands for $\sqrt{d\gs^2}\;V^{\ga} \pa_{\ga}X^{\mu}$. These are D scalar fields on $\gS$ with density weight one half.}
\begin{equation}\label{Variation of S with respect to X (1)}
\gd S = 2\int_{\gS} d^2\gs \left(\pa_{\ga}\gd X_{\mu}\right)  V^{\ga} (VX)^{\mu}
\end{equation}
and integrate by parts to obtain boundary term. This is done by rewriting \eqref{Variation of S with respect to X (1)} as
\begin{equation}\label{Variation of S with respect to X (2)}
\gd S = 2\int_{\gS} d (\gd X_{\mu}) \W \eps_{\ga\gb} V^{\ga} (VX)^{\mu} d\gs^{\gb}.
\end{equation}
Then, the integration by part is straightforward
\begin{align} 
\gd S &= 2\int_{\gS} d\left((\gd X_{\mu})\; \eps_{\ga\gb} V^{\ga} (VX)^{\mu} d\gs^{\gb} \right) - (\gd X_{\mu}) d\left( \eps_{\ga\gb} V^{\ga} (VX)^{\mu} d\gs^{\gb} \right), 
\end{align}
and we can extract equations of motions for the null string 
 \begin{equation}\label{null string dynamical equations}
\pa_{\ga}\left( V^{\ga} \left(VX\right)^{\mu}\right)=0 \qquad \rightarrow\qquad \pab\left( \frac{1}{e} \pab X^{\mu} \right)=0
\end{equation}
together with a general expression for the boundary term:
\begin{equation}\label{null string boundary term}
\gd S_{boundary} = 2\int_{\pa \gS} (\gd X_{\mu})\; \eps_{\ga\gb} V^{\ga} (VX)^{\mu} d\gs^{\gb}.
\end{equation}
Unfortunately there does not seem to be a set of boundary conditions
which gives an interesting theory of open null strings nor null
strings ending on branes. A contraction of the open string algebra can
be done which has been claimed to describe a tensionless open string
\cite{Bonelli:2003kh,Lindstrom:2003mg}, but it is not clear how to recover
it from appropriate boundary conditions on the null string.

Therefore we continue we closed null strings. A clean way to
understand the above integrands is as follows. Start with
\begin{equation}
 (VX)^{\mu} V^{\ga} \; d^2\gs \otimes \pa_{\ga} \in
  \gO^2\left({\gS},T\gS\right),
\end{equation}
which are $D$ vector-valued two-forms on $\gS$, contracting this
object with itself we obtain a 1-form on $\gS$. The resulting form is
just the integrand of \eqref{null string boundary term}:
$\eps_{\ga\gb} V^{\ga} (VX)^{\mu} d\gs^{\gb} \in \gO^1\left(\gS
\right)$. The field equations \eqref{null string dynamical equations}
just state that this form is closed.

Finally, considering variations of the action with respect to an infinitesimal variation of $V$, we get two constraints:
\begin{equation}\label{null string constraints}
V^{\gb} \pa_{\gb}X\cdot \pa_{\ga}X = 0 \quad  \forall \ga \in 0,1 \qquad \rightarrow \qquad \pab X\cdot\pab X=0,\quad \pab X \cdot \pa X = 0.
\end{equation}
These can be directly obtain by varying $V$ in \eqref{eq:LST} or by using the parametrization \eqref{Second Order ambitwistor string} and considering variation of $V$ as
\begin{equation}\label{eq: V variation}
\gd V =  \gd \mu \; \left( \frac{dz d\zb}{e}\right)^{\frac{1}{2}} \otimes
\pa_{z} - \frac{\gd e}{2e} \; \left( \frac{dz d\zb}{e}\right)^{\frac{1}{2}} \otimes
\pab_{\zb}.
\end{equation}
Here $\gd\mu$ is an infinitesimal variation of the almost complex structure $\gd \pab_{\zb} = \gd\mu\; \pa_{z}$.

Altogether, the constraints \eqref{null string constraints} are the
usual null string statement that the pullback of the space-time metric
on the worldsheet
$g_{\ga \gb}=\pa_{\ga}X^\mu\pa_{\gb}X^\nu G_{\mu\nu}$ is degenerate,
with the degeneracy direction given by $V$. Accordingly, integral
lines of $(VX)^{\mu}$ in space-time are null lines and these null
lines are orthogonal to each other.

\subsection{Symmetries of the complexified null string action}

From now on, we also consider the worldsheet variables to be
complex. Accordingly $\gS_{\C}$ is now taken to be a two dimensional
\emph{complex} manifold with holomorphic coordinates $z$ and $\zt$. In
particular, $\zt$ is the complex conjugate of $z$ anymore. Imposing
$\zt= \zb$ amounts to an embedding $\gS \hookrightarrow \gS_{\C}$ of a
usual (one dimensional complex) worldsheet $\gS$ into the complexified
one. The interest of this procedure, of course, lies in the fact that
$\zt= \zb$ is not the only possible embedding, and we intend to make precise in a following work how the ambitwistor string can be seen as an alternative embedding of the null string. When referring to
antiholomorphic functions we will mean holomorphic functions of $\zt$
unless explicitly stated otherwise. All fields are now holomorphic in
$(z, \zt)$ and the worldsheet integral should be seen as a holomorphic
two-form that must be integrated over a two-cycle. In particular, a
choice of real worldsheet gives such a two-cycle. 

\paragraph{Diffeomorphisms}\mbox{}\\
We now consider the action of holomorphic transformations on $\gS_{\C}$ 
\begin{equation}
\left(z ,\zt \right) \mapsto \left(f\left(z,\zt\right) , g\left(z,\zt\right) \right).
\end{equation}
We will refer to these transformations as diffeomorphisms of the complexified worldsheet. Infinitesimal diffeomorphisms are 
\begin{equation}
\left(z ,\zt \right) \mapsto \left(z + \eps\left(z,\zt\right) , \zt + \epst\left(z, \zt\right) \right).
\end{equation}
and can be thought of as the vector field $\bdv = \eps \pa_z + \epst \pat_{\zt}$ on $\gS_{\C}$. These infinitesimal diffeomorphisms act on the fields as
\begin{align}\label{Complex null string Symmetry (1)}
\begin{array}{l}
\mathcal{L}_{\bdv}X = \eps \;\pa_{z} X + \epst \;\pat_{\zb} X,   \\
\mathcal{L}_{\bdv}V=   \left( \frac{dz d\zt}{e}\right)^{\frac{1}{2}}
  \otimes 
\left( -\frac{1}{2e} \left(  \eps \pa e -e\pa \eps + \pat_{\zt}\left( \epst e \right) \right) 
\pab_{\zt}- \left(\pat_{\zt} \eps \right)
\pa_{z} \right).
\end{array}
\end{align}

The Noether current for infinitesimal worldsheet diffeomorphisms is obtained by taking the integrand of the boundary term \eqref{null string boundary term} with $\gd X = \mathcal{L}_{\bdv}X$:
\begin{equation}
J(\bdv) = v^{\ga} T_{\ga\gb} d\gs^{\gb} = \eps \;\frac{1}{e} \pa
X\cdot \pab X \;dz + \epsb \;\frac{1}{e} \pab X\cdot \pab X
\;dz\label{eq:J-diffeos}
\end{equation}
where the energy momentum tensor $T$ is
\begin{equation}\label{null string Energy Momentum tensor}
T= (\pa_{\ga}X)\cdot(VX) V^{\gc}\eps_{\gc\gb} \;d\gs^{\ga}\otimes d\gs^{\gb} = \frac{1}{e} \pa X\cdot \pab X \;dz \otimes dz + \frac{1}{e} \pab X\cdot \pab X \;d\zb \otimes dz.
\end{equation}
As expected, vanishing of the energy-momentum tensor is equivalent to the vanishing of the constraints \eqref{null string constraints}.

Note that the left part of the energy-momentum tensor\footnote{Here $\iota_v$ stands for the interior derivative.} $\iota_{\pa}T= \frac{1}{e} \pa X\cdot \pab X \;dz$ is not simply related to the right part $\iota_{\pat}T= \frac{1}{e} \pab X\cdot \pab X \; dz$. This is in contrast to the Poliakov string where one  left and right movers contributions to the stress energy tensor are related by complex conjugation. This chirality of the null string can be traced back to the fact that $V$ transforms differently under right (i.e $\eps\neq0 , \epst=0$) and left diffeomorphisms (i.e $\eps=0 , \epst\neq0$).

\paragraph{The extra null ray symmetry of the complex null string} \mbox{}\\
The complex null string seems to enjoy one further local
symmetry. Recasting it in the first order form \eqref{Second Order
  ambitwistor string}, this symmetry corresponds to translations along
null geodesics as discussed in \cite{Mason:2013sva}. This symmetry is
also the origin of the interpretation of that model as living on
ambitwistor space, since if we consider target space as parametrized
by the fields $\{P,X\}$, then this extra symmetry implements the
symplectic reduction by the constrain $P^2=0$. This reduced space is
the space of null geodesics, also known as ambitwistor space. The
infinitesimal version of the symmetry can be parametrized by a
$(1,0)$-vector field $\boldsymbol{\ga} = \ga \pa_{z}$ on $\gS$ and
acts on the fields as follows:
\begin{equation}\label{Complex null string Symmetry (2)}
\gd X = \frac{\ga }{e}\;\pab_{\zb} X, \qquad  \gd V = - \frac{\pab_{\zb} \ga}{2e} \; \left( \frac{dz d\zb}{e}\right)^{\frac{1}{2}} \otimes
\pab_{\zb}.
\end{equation}
With the associated Noether Current
\begin{equation}
J(\ga) = \ga \;\frac{1}{e^2} \pab X\cdot \pab X  \;
dz.\label{eq:J-alpha}
\end{equation}
Note that this extra symmetry is not be present in the real case since it does not respect the reality condition $X = \overline{X}$. Even more remarkable is that this symmetry mimics the action of antiholomorphic diffeomorphisms \eqref{Complex null string Symmetry (1)} but is parametrized by a holomorphic vector field. The dictionary between them is as simple as setting
\begin{align}
\epst=\frac{\alpha}{e}. 
\end{align}
It is also easy to see using a Hamiltonian formalism that these two
gauge redundancies are the same on-shell, at least
infinitesimally. This is analogous to what happens in the case of the
Hamiltonian action of the worldline formalism for a massless particle
\cite{Henneaux:1992ig}. There the worldline diffeomorphisms and translations along
null geodesics give the same gauge redundancy of the action
on-shell. It is clear now that to reach the ambitwistor string from
the null string one needs to complexify the latter. This allows us to
access this equivalent parametrization of the antiholomorphic
diffeomorphisms by a holomorphic vector field, and gives a completely
chiral theory, the ambitwistor string.

\subsection{Moduli}
In this section we study the moduli of the complexification of the
vector field $V$. Using the equations for the variation of $V$,
\eqref{Complex null string Symmetry (1)} we define operators $P$ and
$Q$ whose zero modes correspond to automorphisms of the string. Using
the natural pairing, the zero modes of their adjoints $P^\dagger$ and
$Q^\dagger$ are the moduli of the null string. We shall see that after trading the antiholomorphic diffeomorphisms by the holomorphic scaling symmetry from the previous section the results found in \cite{Ohmori:2015sha} for the ambitwistor string are reproduced.

As already explained the moduli $\M$ of the vector field $V$ can be parametrized by a complex structure $d\zb \pa_{\zb}$ and a Beltrami differential $e d \zb \pa_z$, see \eqref{eq:e-transformation}. Then a variation $\gd V$ is an element of the tangent space $T \M$ to the moduli and can be parametrized by a doublet $\left(\gd\mu \; d \zb \pa_z, \gd e \; d \zb \pa_z \right)$ of Beltrami differentials, see \eqref{eq: V variation}.

The infinitesimal gauge transformations of the null string are infinitesimal diffeomorphisms given by $\eps \pa_z$ (left diffeomorphisms) and $\epst \pat_{\zt}$ (right diffeomorphism) \footnote{Recall that $z$ and $\zt$ are considered independent complex variables}. A generic variation of $V$ under infinitesimal diffeomorphisms defines a map $\Gamma\left[T_{\C}\gS\right] \to T\M$ given by two operators
\begin{equation}
\begin{array}{ccccccccc}
P: &\Gamma\left[T^{(1,0)}_{\C}\gS\right] &\to& T\M &,\qquad \qquad  & Q:&
  \Gamma\left[T^{(0,1)}_{\C}\gS\right] &\to& T\M,\label{eq:PQ-def}   \\
  & \eps \pa & \mapsto& \Ld_{\eps \pa} V& & & \epst \pat & \mapsto &  \Ld_{\epst \pat} V
\end{array}
\end{equation}
comparing \eqref{eq: V variation} with \eqref{Complex null string Symmetry (1)} we obtain
\begin{align}
\begin{array}{llllllll}
P\left(\eps\,\pa_{z}\right) &=\left(P_\mu\,,\, P_e\right) &= \left(-\pat_{\zb} \eps \;d\zt \,\pa_z  \,,\, \left(\eps \,\pa_{z} e -e\,\pa_{z} \eps\right)\;d\zt \,\pa_z \right),
\end{array}
\end{align}
\begin{equation}
\begin{array}{llllllll}
Q\left(\epst\,\pab_{\zb}\right) &=\left(Q_\mu, Q_e\right) &= \left( 0 \,,\, -\pat_{\zt} \left(\epst e\right) \; d\zb\, \pa_z\right).
\end{array}
\end{equation}

By construction, $G = Im(P)\cup Im(Q)$ is the subspace of $T\M$
spanned by the gauge transformations. We are looking for variations of
the gauge parameters that cannot be the result of a gauge
transformation. By picking a metric on $T\M$, these non-gauge
variations can be taken to be $G^{\perp}$. One can easily get such a metric by making a choice of hermitian metric on $\gS_{\C}$, $ds^2 = dz \odot d\zb+ d\zt \odot d\bar{\zt}$. If $u$ and $v$ are any tensor of same type, we note $\overline{u}v$ the hermitian pairing induced by the above metric. 

We can define $G^{\perp}$ to be the subset of $T\M$ such that for all $ \eps, \epst $:
\begin{equation}
\begin{aligned}
	\int_{\gS} dz d\zt  \left(\;\gd\bar{\mu}\; P_{\mu}(\eps)  + 
	\gd\bar{ e} P_{e}(\eps) 
	\;\right) &=0, \\
	\int_{\gS} dz d\zt \left(\; \gd \bar{\mu}\; Q_{\mu}(\epst) + 
	\gd\bar{ e} Q_{e}(\epst) 
	\;\right) &=0.
\end{aligned}
\end{equation}
 We are therefore looking for $Ker( P^{\dagger}) \cap Ker(Q^{\dagger})$, with
$P^{\dagger} : T\M \to \Gamma [T^{(1,0)}_{\C}\gS]\ $,
$ Q^{\dagger} : T\M \to \Gamma [T^{0,1}_{\C}\gS] $ defined by
\begin{align}
	\int_{\gS} dz d\zt \left(\;\bar{\gd \mu}\; P_{\mu}(\eps)  + 
	\gd e P_{e}(\eps) 
	\;\right) &=  \int_{\gS} dz d\zt \; \overline{P^{\dagger} (\gd \mu, \gd e)}\; \eps \label{eq:P-dagger}\\
	\int_{\gS} dz d\zt \left(\;\gd\bar{\mu}\; Q_{\mu}(\epst)  + 
	\gd\bar{e} \; Q_{e}(\epst) 
	\;\right) &=  \int_{\gS} dz d\zt \;\overline{Q^{\dagger} (\gd \mu, \gd e)}\; \epst.\label{eq:Q-dagger}
\end{align}
We can obtain these operators explicitly by an integration by parts:
\begin{equation}
\begin{aligned}
	&\int_{\gS} dz d\zt \left(\; \gd\bar{\mu} \; \left( P_{\mu}(\eps) + Q_{\mu}(\epst)\right)  + \gd\bar{e} \; \left( P_{e}(\eps) +  Q_{e}(\epst)\right) \;\right) \\
	=& \int_{\gS} dz d\zt \left( \gd\bar{\mu} \; \pat \eps + \gd\bar{e} \; (\eps\pa e-e\pa\eps) + \gd \bar{e} \; \pat(\epst e)\right) \,, \\
	=& \int_{\gS} dz d\zt \left(\eps\;\left( -\pat \; \gd\bar{ \mu} + \pa e\, \gd\bar{e} + \pa\left(e \; \gd \bar{e} \right)\right) - \left(\epst e\right)\;\pat \gd\bar{e}  \right)\,.
\end{aligned}\label{eq:P-dagger-Q-dagger}
\end{equation}
Requiring that this holds for any $\eps$ and any $\epst$ gives the equations
\begin{align}
 &\pab \gd \bar{\mu} - \pa e\delta \bar{e} - \pa(e\delta \bar{e}) = 0\nonumber\\
 &e\,\pab \gd \bar{e} = 0.
\end{align}
If we now parametrize the diffeomorphisms in $\bar{z}$ by a
holomorphic field $\alpha = e\epst$ using the equivalence of
these diffeomorphisms with the scaling symmetry, we recover the same
equations for the moduli as \cite{Ohmori:2015sha}. Here too it was
essential that the null string be complexified in order to match the
ambitwistor string.

\section{Symmetry algebra}
\label{sec:gca}

In this section we come back on the symmetry algebra of the null
string. Following recent
terminology~\cite{Bagchi:2013bga,Bagchi:2016yyf,Bagchi:2015nca} is
called a 2-dimensional Galilean Conformal Algebra, $\gca_2$. This
algebra is isomorphic to the 3-dimensional Bondi-Metzner-Sachs $\bms_3$
algebra -- the symmetry algebra of the null boundary of 3-dimensional
Minkowski spacetime. This isomorphism is at the root of various
conjectures concerning flat space
holography~\cite{Bagchi:2009my,Barnich:2010eb,Bagchi:2010eg} which
have triggered interest for $\gca_2$ representations and supersymmetric
extensions
thereof~\cite{Bagchi:2009pe,deAzcarraga:2009ch,Sakaguchi:2009de,Bagchi:2009ke,Mandal:2010gx,Barnich:2014kra,Banerjee:2015kcx,Mandal:2016lsa,Mandal:2016wrw,Banerjee:2016nio,Lodato:2016alv,Fuentealba:2017fck,Basu:2017aqn,Oblak:2016eij,Campoleoni:2016vsh}. See
also \cite{Duval:2014uva} for a connection with Carrollian
ultra-relativistic physics.\footnote{It is known that in two
  dimensions, ultra- and non-relativistic physics are classically
  equivalent, essentially because there are as many space and time dimensions.}
In the text, we frequently use the BMS terminology, and call the GCA
transformations superrotations and supertranslations.

Our motivation in studying GCA's comes from wanting to set up a vertex
operator formalism for the null string where the loop-momentum zero
modes are already integrated. In addition, the symmetry algebra of the
null string is a GCA and not just the traditional Virasoro algebra of
string theory, it would appear necessary to start from scratch and
work out the equivalent of the basic tools that we have in ordinary
CFTs; state-operator map and vertex operator formalism.

In this section, we will show that the representation theory of the
$\gca_2$ for the null string actually forces the representations to
truncate down to the usual Virasoro representations. To do so, we will
mostly use of the analysis of null states of the $\gca_2$
presented in \cite{Bagchi:2009pe}.

This surprising fact justifies intuitively why it has been possible to
use standard CFT tools so far both in the ambitwistor and null string
(see in particular the recent work on one-loop null string amplitudes
of \cite{Yu:2017bpw}).
It will also shed a new light on the remarkable chirality of the
ambitwistor and null strings and the truncation of their spectrum.

Before starting, we would like to briefly comment on conformal
non-relativistic symmetries, mostly to disambiguate the
terminology. The algebra studied here is not the Schr\"odinger algebra
of \cite{Hagen:1972pd,Niederer:1974ba,Henkel:1993sg,Nishida:2007pj}
but the algebra obtained by an In\"on\"u-Wigner contraction of the
usual Poincar\'e algebra. It exists in any dimensions, and only in two
dimensions it has the infinite dimensional extension which also makes
it a contraction of a product of two Virasoro
algebras~\cite{Henkel:2006tg,Hosseiny:2009jj,Bagchi:2009pe}. We refer
to \cite{Duval:2009vt,Bagchi:2009pe} for further details and
references.

\subsection{Gauge fixing and residual symmetries}
Consider partially gauge-fixing the null string action \eqref{eq:LST}
by making a choice of complex structure. Looking at the variations of
$V$, \eqref{Complex null string Symmetry (1)}, we see that the
``right-diffeomorphisms'' ($\eps=0$) preserve this gauge
choice. However, requiring that ``left-diffeomorphisms'' ($\epst=0$)
preserve this complex structure imposes $\tilde{\partial} \eps =0$,
that is, it is only a function of $z$.
\begin{align}
\begin{array}{ll}
\begin{array}{ll}
\mathcal{L}_{\boldsymbol{\eps} } X = \eps \;\pa_{z} X,   \\
\mathcal{L}_{\boldsymbol{\eps} } V = \left( \frac{dz d\zt}{e}\right)^{\frac{1}{2}} \otimes \left( -\frac{1}{2e} \left(  \eps \pa e -e\pa \eps ) \right) \;
\pat_{\zt} \right),
\end{array} &\qquad
\begin{array}{ll}
\mathcal{L}_{\boldsymbol{\epst} } X = \epst\;\pat_{\zt} X, \\
\mathcal{L}_{\boldsymbol{\epst} } V =\left( \frac{dz d\zt}{e}\right)^{\frac{1}{2}} \otimes - \frac{\pat_{\zt}\left(\epst e \right)}{2e} 
\;\pat_{\zt}.
\end{array}
\end{array}
\end{align}
We can further gauge-fix by choosing a particular value for the Lagrange multiplier scale field $e$. Residual symmetries then have to satisfy
\begin{align}
\eps \pa_{z} e -e\pa_{z} \eps + \pat_{\zt}( \epst e ) = 0,& \qquad  \pat_{\zt} \eps =0.  
\end{align}
Taking $e$ to be constant, these symmetries are generated by vector fields of the form
\begin{equation}\label{GCA vector fields}
  \epsilon \pa_z + \tilde \epsilon \pat_{\zt} = 
f(z) \pa_{z} + \left(\zt \pa_{z} f(z) + g(z)\right) \pat_{\zt} 
\end{equation}
where $f,g$ are any holomorphic functions. The associated Noether currents are
\begin{equation}
J_f = f(z) \left( \pat X . \pa X  - \zt \pa \left(\pat X. \pat X\right)\right)d\zt
\qquad \text{and} \qquad \tilde{J}_{g}= g(z) \left( \pat X . \pat X
\right)d\zt.\label{eq:noether-GCA}
\end{equation}
These vector fields form a GCA which play the same role in in the null string as the Virasoro algebra does in the usual string. Note that the GCA contains a single copy of the Virasoro algebra as a subalgebra giving the null string its chiral character. 

Put differently, the transformation~(\ref{GCA vector fields}) defines
two operators that we can call $L(F)$ and $M(g)$ whose mode expansion
are given by
\begin{equation}
  L(f)=\sum_{n\in \mathbb{Z}}f_n L_n,\quad M(g)=\sum_{n\in
  \mathbb{Z}} M_n g_n
\label{eq:mode-exp-L-M}
\end{equation}
with 
\begin{align}
  L_n = -z^{n}(z\partial_z + (n+1)\zt \partial_{\zt})\,,\qquad
M_n= z^{n+1}\partial_{\zt}\,.
\label{eq:GCA-rep}
\end{align}
In the BMS language, $L_n$ and $M_n$ are the generators of
superrotations and supertranslations, respectively.\footnote{The
  combination
  $ L_n' = L_n-i(n+1) \frac \zt z M_n = -z^{n+1}\partial_z $ generates
  exactly chiral conformal transformations we are after. However,
  since the change of generators involves the variables themselves, it
  is not clear what can be made of this observation.}
They obey the
following commutation relations
\begin{equation}
  \label{eq:gca-classical}
  [L_n,L_m]=(n-m)L_{n+m},\quad [L_n,M_m]=(n-m)M_{n+m},\quad [M_n,M_m]=0\,.
\end{equation}
At the quantum level, central extensions are admissible. The centrally-extended algebra is
\begin{equation}
  \begin{aligned}
    \label{eq:gca-quantum}
    [L_n,L_m]=&(n-m)L_{n+m} + \frac{c_L}{12}m(m^2-1)\delta_{{m+n},0}\\ 
    [L_n,M_m]=&(n-m)M_{n+m}+ \frac{c_M}{12}m(m^2-1)\delta_{{m+n},0}\\ 
    [M_n,M_m]=&0\,.
  \end{aligned}
\end{equation}
For the ambitwistor string, $c_M=0$ and $c_L=d-2$ is canceled by the
inclusion of the $b$-$c$ and $\tilde b$-$\tilde c$ ghost systems.  The
vacuum chosen to study the representations of the GCA is the same as
the one used in the ambitwistor quantization and is defined
by\footnote{In the other quantization, supposed to produce a
  higher-spin theory~\cite{Gamboa:1989zc,Lizzi:1986nv}, the operator
  ordering stipulates that all the modes of $P$ annihilate the
  vacuum. Therefore, all the modes of the constraint annihilate the
  vacuum $\forall n\in \mathbb{Z},\,L_n|0\rangle=M_n|0\rangle=0,$ and
  it is not clear how to build non-trivial representations. This may
  reflect that the theory is likely to be free, as expected from the
  Coleman-Mandula theorem.}
\begin{equation}
  \label{eq:GCA-vacuum}
  L_{n}|0\rangle=0\,,\quad   M_{n}|0\rangle = 0\,, \quad \forall n\geq0\,.
\end{equation}

\subsection{GCA Hilbert space and null states}
\label{sec:gca-hilbert-space}
We now proceed to investigate the GCA representations. We will use the
analysis of~\cite{Bagchi:2009pe} and argue that they simply truncate
down to a chiral Virasoro representation.

The upshot is that due to how the $P^2=0$ constraint is imposed,
the GCA action automatically descends to a chiral CFT action at the
level of the spectrum.

We would like to conjecture that for this reason we can have a
well-defined state-operator map for the chiral CFT as well as a
standard vertex operator formalism. It still remains an important
question to understand these issues in full generality in the GCA and
may open the way towards massive theories for instance, where the
constraint $P^2=0$ should not be applied.

We start by reviewing some elements of the analysis of
\cite{Bagchi:2009pe} on the representations of the $\gca_2$
algebra. We look at states with well-defined scaling properties
\begin{equation}
L_0 |\Delta\rangle=\Delta|\Delta\rangle.\label{eq:scaling}
\end{equation}
Then, since $[L_0,M_0]=0$ the representations are actually indexed by another quantum number $\xi$ called ``rapidity''~\cite{Bagchi:2009ca,Bagchi:2009pe}
\begin{equation}
L_0 |\Delta,\xi\rangle=\Delta|\Delta,\xi\rangle,\quad M_0 |\Delta,\xi\rangle=\xi|\Delta,\xi\rangle.\label{eq:rapid}
\end{equation}
Descendant states are then built out by the successive action of the
operators $L_{-n},M_{-m}$, $n,m>0$.

We now follow the analysis of \cite[sec 5]{Bagchi:2009pe} on the GCA
null states. Here $c_M=0$ and the physical state conditions impose
$\Delta=2$, and, importantly, $\xi=0$. The first condition states that
physical states are primaries of conformal weight two. The second
condition is on-shellness of the state, i.e.  $k^\mu k_\mu=0$ for a
state with momentum $k^\mu$.

This is this last condition that actually implies that the null string
does not use of the full GCA symmetry. We will see that it implies
that the $M_{-n}$ descendants decouple. The argument adapted from
\cite{Bagchi:2009pe}, goes as follows.

At level one, there are two descendant states
$L_{-1} |\Delta,0\rangle$ and $M_{-1} |\Delta,0\rangle$.  It is
immediate to see that the second one, $M_{-1} |\Delta,0\rangle$, is
orthogonal to all other states in the Hilbert space. Therefore
$M_{-1} |\Delta,0\rangle=0$.  At level two, descendants made of powers
of $M_{-1}$ and $M_{-2}$ are the following states
\begin{equation}
  \label{eq:level-2}
  (M_{-1})^2  |\Delta,0\rangle,\quad  L_{-1}
  M_{-1}|\Delta,0\rangle\,,\quad M_{-2} |\Delta,0\rangle\,.
\end{equation}
The first two states vanish immediately, because
$M_{-1} |\Delta,0\rangle=0$. The second state is, again, orthogonal to
all other states, precisely because $M_0 |\Delta,0\rangle=0$.  The
whole sector of the Hilbert space made of $M_{-n}$'s is therefore null
and decouples from the physical Hilbert space. We are then left with a
chiral Virasoro module.  This is the reason why it is possible to treat
the null string and ambitwistor string as a chiral CFT, and
intuitively, is the origin of the holomorphicity of all twistor string
models.

\subsection{Gauge-fixing the global GCA}

After the gauge-fixing, there is still a residual gauge symmetry which
is given by the global part of the GCA. Below we explain how this
residual gauge redundancy is removed by fixing the positions of 3
operators, in analogy with the similar string-theoretic version.

The method previously used in \cite{Bagchi:2009pe,Bagchi:2009ca} was
to consider the $\gca_2$ as a contraction of the usual
$\mathrm{Vir}\times \overline{\mathrm{Vir}}$ algebra, under which the
coordinates $z,\tilde z$ are scaled as
\begin{equation}
  \label{eq:squeeze}
  \begin{aligned}
    z=t+\epsilon x\\
    \zt = t-\epsilon x
  \end{aligned}
\end{equation}
with $\epsilon\to0$. An $SL(2,\mathbb{C})$ transformation then induces the following transformation
\begin{equation}
  \label{eq:SL2-squeeze}
  t+\epsilon x \to \frac{a(t+\epsilon x)+b}{c(t+\epsilon x)+d}=
  \frac{at+b}{c t+d}+\epsilon \frac{x}{(c t+d)^2}\,.
\end{equation}

Here, again following our wish to work out the details of the model,
we will derive these relations from the explicit form of the global
GCA transformations.

We start from the representation of eq.~(\ref{eq:GCA-rep}). The
generators $L_0,L_1,L_{-1}$ and $M_0,M_1,M_{-1}$ constitute the global
part of the gauge group. Their expressions read
\begin{equation}
  \label{eq:global-naive}
  \begin{tabu}{lll}
    L_{-1}=-\partial_t\,,\quad& L_0=x \partial x-t\partial_t\,,&
    L_1=-2tx \partial_x -t^2 \partial_t\,,\\
    M_{-1}=\partial_x\,,& M_0=t\partial_x\,,& M_1= t^2 \partial_x\,.
  \end{tabu}
\end{equation}

We claimed that these generators are globally defined, but there is a
subtlety here. Due to the term $-2tx \partial_x$, $L_1$ is not well
defined for $t\to \infty$, unless $x=0$. We shall see later that it is
always possible to fix $x=0$, and moreover that these terms produce
only off-diagonal terms in the determinant of the zero modes which
anyway do not contribute to the total determinant.  It is also
intriguing to see that, at fixed $t$, all the $M_n$'s for all
$n\in \mathbb{Z}$ are well defined, but only $M_{-1}$ is for all
values of $t$. Infinitesimal transformations associated to these six
generators can be written easily, an read for the $L_{-1},L_0,L_1$
with parameters $\delta a_{-1},\delta a_0,\delta a_1$:
\begin{equation}
  \label{eq:L-infinitesimal}
  \begin{aligned}
    \delta t & = \delta a_{-1} + (\delta a_0) t - (\delta a_1) t^2\\
    \delta x &= -\delta a_0 x- 2(\delta a_1) t x
  \end{aligned}
\end{equation}
and for the $M_i$'s with parameters $ \delta b_{-1},\delta b_0,\delta b_1$
\begin{equation}
  \label{eq:L-infinitesimal2}
  \begin{aligned}
    \delta t & =0\\
    \delta x &= \delta b_{-1} +(\delta b_0) t +(\delta b_1) t^2
  \end{aligned}
\end{equation}
To integrate to the finite form, in principle one has to solve a
differential equation. Take the special conformal transformation of
the conformal group, generated by $\delta z = -(\delta \alpha) z^2$. It is
solved by writing $\frac{\delta z}{z(\alpha)^2} = \delta \alpha$
which gives $1/\tilde{z}-1/z = \alpha$, i.e.
$\tilde{z} = \frac{z}{1+\alpha z}$. In the case of the GCA
transformations, only the $L_1$ requires a little care. Calling
$s=a_1$, it reads
\begin{equation}
\begin{aligned}
  \frac{\delta t}{t(s)^2}=-\delta c,\quad \frac{\delta
    x}{x(s)}=-2\delta s \times t(s)
\end{aligned}\label{eq:L1finite}
\end{equation}
where we have made the dependence on $c$ explicit in the functions
$t,x$. Integrating $t$ gives $t(s)=t(0)/(1+s t(0))$, which can be
plugged into $\delta x/x$ to give $x(s) = x(0)/(1+s t)^2$. Combining
with $L_0$ and $L_{-1}$ we obtain the following finite
transformations:
\begin{equation}
  t\to \tilde{t} = \frac{a t+b}{c t+d},\qquad x\to
  \tilde{x}=\frac{x}{(c t+d)^2}
\label{eq:finte-L}
\end{equation}
for the $L_i$'s and
\begin{equation}
  t\to \tilde{t} =t,\qquad x\to
  \tilde{x}=x+e+f t+ g t^2
\label{eq:finite-M}
\end{equation}
for the $M_i$'s.

Given 3 points $(t_i,x_i)$ on $\mathbb{C}^2$ we apply the finite transformations above to perform the usual gauge fixing of the $t$'s
to $0,1,\infty$ and fix $x_1,x_2,x_3$ to zero. For four points, we have determined explicitly that this produces the
two GCA-independent quantities found in \cite{Bagchi:2009pe} using the
previously described squeeze limit:
\begin{equation}
  \label{eq:gca-cross-ratios}
  t=\frac{t_{23}t_{14}}{t_{12}t_{34}}\,,\qquad 
  \frac x t = \frac{x_{12}}{t_{12}} - \frac{x_{14}}{t_{14}} - \frac{x_{23}}{t_{23}} + \frac{x_{34}}{t_{34}}\,.
\end{equation}
This means that for $x_1=x_2=x_3=0, t_1=0,t_2=1$ and $t_3=\infty$, we
just have $t_4=t$ and $x_4=x$. 
The finite $BMS_3$ transformations have been computed in
\cite{Barnich:2016lyg}, it would be interesting to understand if they
have any geometrical interpretation in the $\gca_2$ side.

Lastly we compute the Jacobian for gauge-fixing the global GCA. In a
BRST framework this comes from integrating out the zero modes of the
ghosts associated to the constraints \eqref{null string
  constraints}. There are six ghosts, one for each global generator of
the GCA \eqref{eq:global-naive}. Therefore there are six global
sections which we can fix by picking three points on the worldsheet
$\{(t_1,x_1),(t_3,x_3),(t_3,x_3)\}$ and calculating the determinant of
the matrix of zero mode sections evaluated at these points
\begin{equation}
  \label{eq:c-determinant}
  M=
  \begin{pmatrix}
    A & 0 \\ B & -A  
  \end{pmatrix},  \mathrm{where}~
  A=
  \begin{pmatrix}
    1 & 1 & 1 \\
    t_1 & t_2 & t_3\\
    {t_1}^2 & {t_2}^2 & {t_3}^2
  \end{pmatrix},\,\,
  B=
  \begin{pmatrix}
    0 & 0 & 0 \\
    x_1 & x_2 & x_3\\
    -2x_1{t_1} & {-2 x_2 t_2} & {-2 x_3t_3}
  \end{pmatrix}\,.
\end{equation}
This matrix has an off-diagonal part because the $L_n$ and $M_n$
generators do not commute, in contrast to the left- and right-handed
Virasoro algebra in usual 2D CFTs. However, the off-diagonal does not
contribute to the determinant which is
\begin{equation}
  \label{eq:det-c}
  \det(M) =- ((t_{1}-t_2)(t_{2}-t_3)(t_{3}-t_1))^2\,.
\end{equation}
and is precisely the same as found in the ambitwistor string. Note
that since the $x$ coordinates decouple from the determinant their
fixed values are immaterial to the correlation function, effectively
all that is needed to fix the global GCA is choosing three points in
the $t$ coordinate. This is precisely what happens in the ambitwistor
string, where one only fixes three holomorphic coordinates to fix the
global GCA at tree-level.

\section{Operator formalism and scattering equations}
\label{sec:op}

\subsection{Formalism}
\label{sec:formalism}

As we mentioned, the ambitwistor complexified gauge is an elegant way to
reproduce the CHY formulae. However, subtleties show up at loop-level
which render this power somewhat useless, in particular when
discussing questions related to modular invariance and the role of the
loop momentum for instance. In this section, we set up an operator
formalism\footnote{Here done somewhat heuristically since we neglect
  the ghosts for the most part.} which will remain somewhat agnostic
about the complexification since the manipulations are purely
algebraic. We then use it to gain insights into the appearance of the
scattering equations in the ambitwistor string by comparing the
amplitude computed in these two different ways. We also make
connection with an interesting one-loop computation using CFT methods
presented in \cite{Yu:2017bpw}.

The formalism will essentially follow the analogous operator
construction in string theory, presented in the classic
reference~\cite{Green:1987sp}.  To set up the formalism, we consider the
canonical quantization of the null string in Schild's
gauge~\cite{Gamboa:1989zc}:
\begin{equation}
V\sim \partial_\tau\,.\label{eq:schildsgauge}
\end{equation}
Note that this is a Lorentzian gauge fixing condition and should be
contrasted with the more Euclidean condition chosen earlier
in~(\ref{eq:V-dbar}).

Even though the amplitude calculation is only well-defined in the
model with two supersymmetries, we work in the purely bosonic model
since it has all the important features without the added
combinatorial complexity of having the fermions. This feature will
prove sufficient to exhibit the essential properties of the model, the
scattering equations in particular.

The relevant field to quantize is $X$, for which the equation of
motion $\partial_\tau^2X=0$ gives the following classical solutions
\begin{align}
 X(\tau,\sigma)=Y(\sigma)+\tau P(\sigma)
\end{align}
which we expand in modes
\begin{align}
 Y(\sigma)=\sum_{n\in\mathbb{Z}}y_n e^{-i\sigma n},\;\;\;\;\;\; P(\sigma) = \sum_{n\in\mathbb{Z}}p_n e^{-i\sigma n}\,,
\end{align}
with canonical commutation relations $[y_n,p_m]=i\delta_{m+n,0}$. Here
and below we omit Lorentz indices for convenience. In this gauge, the
two constraints are given by
$\partial_\tau X\cdot \partial_\sigma X=0$ and
$\partial_\tau X\cdot \partial_\tau X=0$. The mode expansion of these
in terms of the corresponding $L$ and $M$ generators and their
commutation relations can be found in
\cite{Bagchi:2015nca,Casali:2016atr}. For what follows, we only need
the zero modes of these operators: $L_0$ generates rotations along the
circle and $M_0$ is the worldsheet Hamiltonian (we provide their
explicit expression below).

We postulate that a vertex operator with momentum $k$ placed at the
$(\sigma,\tau)=(0,0)$ assumes the following form
\begin{align}
 V(0,0):=(\epsilon\cdot\dot{X}(0,0))^2e^{ik\cdot X(0,0)}=(\epsilon\cdot P(0))^2e^{ik\cdot Y(0)}.
\end{align}
where $\varepsilon_{\mu\nu}=\epsilon_{(\mu} \epsilon_{\nu)}$ is the
graviton's polarization.  The amplitude is obtained from a correlator
of local insertions of these operators. First we apply a vertex
operator to the incoming vacuum, propagate this state using the
worldsheet propagator $\Delta$, act with another vertex operator, and
so on, until we contract with the outgoing vacuum. That is, at four
points,
\begin{align}\label{corre}
 \langle \epsilon_1;k_1 | V_2(0,0)\Delta V_3(0,0)|\epsilon_4;k_4\rangle.
\end{align}
The full amplitude is obtained by summing over permutations of the
external particles. We use the following expression for the worldsheet
propagator;
\begin{align}
 \Delta=\frac{\delta(L_0-2)}{M_0}=\int d\rho d\phi e^{-\rho M_0}e^{-i\phi(L_0-2)}.
\end{align}
This formulation is closely related to one used in \cite{Li:2017emw}
for the HSZ string and has its origin in the descent procedure from
\cite{Adamo:2013tsa}.  It would be interesting to compare this
expression with the expression derived rigorously in
\cite{Ohmori:2015sha}.

The zero point energy contribution for $L_0$ occurs when one picks the
ambitwistor vacuum~(\ref{eq:GCA-vacuum}), also defined in terms of the
$p_n,y_m$ modes:
\begin{equation}
p_{n}|0\rangle=0;\;\;\;y_{n}|0\rangle=0\;\;\forall n>0,
\end{equation}
and equivalently the following operator ordering
\begin{equation}
  \label{eq:normal-ordering}
  :y_n p_m: =
  \begin{cases}
    y_n p_m \mathrm{~if~} m>0\\
    p_m y_n \mathrm{~if~} n>0
  \end{cases}
\end{equation}
which is the appropriate one here. These operators are responsible for
moving vertex operators along the worldsheet as
\begin{align}
 e^{-\rho M_0}e^{2i\pi\phi L_0}V(0,0)e^{\rho M_0}e^{i\phi L_0}=V(\rho,\phi)
\end{align}
and are given by
\begin{align}
 L_0=\sum_{n\in\mathbb{Z}}n:p_{-n}\cdot y_n:\,,\quad M_0=\frac{1}{2}\sum_{n\in\mathbb{Z}}p_{-n}\cdot p_n.
\end{align}
The correlator \eqref{corre} becomes
\begin{align}\label{corre2}
 \int d\rho d\phi\langle \epsilon_1;k_1 | V_2(0,0)V_3(\rho,\phi)|\epsilon_4;k_4\rangle
\end{align}
The only place where $Y$ appear is in the exponentials, so the
commutator between them and polynomials of $P$ are easy to evaluate
and will not have any dependence on $\rho$. The only term with non-trivial dependence on the modulus $\rho$
is given by commuting the exponential parts of $V_2$ through the other
vertex operators, for example
\begin{align}
e^{ik_2\cdot Y_-(0)} e^{ik_3\cdot Y_+(\phi)+i\rho k\cdot P_+(\phi)}=e^{ik_3\cdot Y_+(\phi)+i\rho k\cdot P_+(\phi)}e^{ik_2\cdot Y_-(0)} e^{-i\rho k_2\cdot k_3G(0,\phi)}
\end{align}
where the $Y_{\pm}(\phi)=\sum_{\pm n\ge 0}y_n e^{-in\phi}$ and the
same for $P$. The function
$G(\phi_1,\phi_2)=(1-e^{-i(\phi_1-\phi_2)})^{-1}$ is the propagator on
the cylinder. We give more details on its computation in the next
section.

The full computation of the correlator for an $n$ point scattering is
actually done using the Baker-Campbell-Hausdorff formula. Its full
dependence of on the moduli $\rho$ comes in the exponential
\begin{align}\label{eq: Op scattering eq}
 \exp\left(i\rho\left(k_3\cdot k_1+\frac{k_3\cdot k_2}{1-\frac{1}{z}}\right)\right)
\end{align}
with $z=e^{-i\phi}$. Here is where the complexification comes
in. Complexifying the moduli and changing the integration contour of
$\rho$\footnote{Together with a change of variable
  $\rho\rightarrow\frac{\rho}{z}$} such that the above exponential
integrates to a delta-function, its
argument coincides with the four point scattering equation
\begin{align}
 k_3\cdot P(z) = \sum_{i\neq3}\frac{k_3\cdot k_i}{z-z_i}=0.
\end{align}
Here the gauge $\{z_1,z_2,z_4\}=\{0,1,\infty\}$ appears naturally. In
the original coordinates this corresponds to picking
$\{\sigma_1,\sigma_2,\sigma_4\}=\{i\infty,0,-i\infty\}$, which can
only be achieved with complex moduli.

The inclusion of fermions does not change the above calculation of the
the exponential factors, the same is true if more vertex operators are
included. The dependence on the moduli associated to the Hamiltonian
$M_0$ is always exponential and, by picking the right contour, can be
integrated into the delta functions imposing the scattering
equations. This way of obtaining the scattering equations is
reminiscent of the descent procedure described in
\cite{Adamo:2013tsa}, but here we made no use of the CFT
description. To recover actual gravity amplitudes we simply use the
$\mathcal{N}=2$ version of the null string and consider correlators of
the form
\begin{align}
 \langle \epsilon_1;k_1|V_2\Delta V_3\Delta \cdots \Delta V_{n-1}|\epsilon_n; k_n\rangle
\end{align}
and sum over permutations. The vertex operators have the form
\begin{align}
 V(0,0)=(\epsilon\cdot P +\epsilon\cdot\psi k\cdot\psi)^2 e^{ik\cdot Y}(0,0).
\end{align}
After expressing all the propagators in terms of moduli and commuting them through to the vacuum the calculation is essentially the same as in the ambitwistor string up to change of coordinates in the moduli space.

\subsection{Cylinder propagator and n-point scattering equations}

\label{sec:cylind-prop-n}

Here we give more details on the computation of the propagator
$\langle X X\rangle$ on the cylinder using the operator formulation. A
similar computation was performed proposed in \cite{Yu:2017bpw} using
a operator and path integral methods -- we find agreement with these
results. With this propagator we see how the scattering equations in
the operator formalism arise from a contour deformation of the time
variable~$\tau$. Similar observations were made in
\cite{Siegel:2015axg,Yu:2017bpw}. It is important for us to revisit
these analyses because it allows us to constrain further the
complexification of the null string.
Using the definitions of the previous section, the correlator is given
by:
\begin{equation}
  \label{eq:corr-XX-def}
\langle X(\tau_1,\sigma_1) X(\tau_2,\sigma_2)\rangle = T\left( X(\tau_1,\sigma_1) X(\tau_2,\sigma_2)\right)-: X(\tau_1,\sigma_1) X(\tau_2,\sigma_2):
\end{equation}
where $T(\ldots)$ and $:\ldots:$ denote time and normal ordering,
respectively. The usual ordering would be $\tau$-ordering,
however the computation does not change if we use a
$\sigma$-ordering. The reason why we make this comment is because
there is an intuitive sense in which the ambitwistor normal ordering
amounts to exchanging space and time on the worldsheet, as described
by Siegel in \cite{Siegel:2015axg}.

Suppose {$\tau_1>\tau_2$}, (or $\sigma_1> \sigma_2$):
\begin{equation}
  \begin{aligned}
    \langle X(\tau_1,\sigma_1)X(\tau_2,\sigma_2)\rangle&
=\sum_{n,m\in\mathbb{Z}} \big((y_n +\tau_1 p_n)( (y_m +\tau_2 p_m)
-: (y_n +\tau_1 p_n)( (y_m +\tau_2 p_m):\\
&=\sum_{n>0,m<0}\left(\tau_1(p_n y_{m}-y_{m}p_n)+\tau_2(y_n p_{m}-p_{m}y_n)\right)
\big)e^{i n
  \sigma_1 +i m \sigma_2}\\
&= -i (\tau_1-\tau_2)\sum_{n>0}e^{in(\sigma_1-\sigma_2)}
  \end{aligned}
\end{equation}
finally giving
\begin{equation}
  \label{eq:corr-XX-asym}
  \langle X(\tau_1,\sigma_1) X(\tau_2,\sigma_2)\rangle =-i
  (\tau_1-\tau_2)\frac{z_1}{z_1-z_2}
\end{equation}
where we put $z_i=\exp(i \sigma_i)$. In terms of $\sigma$ and $\tau$
this can be rewritten
$\langle X(\tau_1,\sigma_1) X(\tau_2,\sigma_2)\rangle =
(\tau_1-\tau_2) \left(
  \cot\left(\frac{\sigma_1-\sigma_2}{2}\right)+1\right)/2$ where the
invariance by translation symmetry is now obvious.  The constant piece
will drop out of the propagator by $1\leftrightarrow2$ symmetry, so we
can as well remove it from the start. This amount to replace the
previously derived propagator by
\begin{equation}
  \label{eq:corr-XX-sym}
  \langle X(\tau_1,\sigma_1) X(\tau_2,\sigma_2)\rangle =-\frac i2
  (\tau_1-\tau_2)\frac{z_1+z_2}{z_1-z_2}
\end{equation}

The null-string's Koba-Nielsen factor, abbreviated $\sum k_i\cdot k_j
\langle X_i X_j\rangle$, then reduces to
\begin{equation}
  \label{eq:null-KN}
  \begin{aligned}
 \sum k_i\cdot k_j \langle X_i X_j \rangle& =-\frac i2 \sum_{i,j} k_i\cdot
 k_j\tau_{ij} \frac{z_i+z_j}{z_i-z_j}\\
 &= -\frac i4\sum_{i<j} k_i \cdot k_j \tau_i \frac{z_i+z_j}{z_i-z_j}\\
&= -\frac i2\sum_{i=1}^n
\tau_i z_i\,\left(\sum_{j=1}^n 
  k_i \cdot k_j  \frac{1+z_i/z_j}{z_i-z_j}\right)
  \end{aligned}
\end{equation}
where to go from the first to second line we used momentum conservation.

Then, as argued above, the $\tau_i$ integration should be 
complexified in such a way as to give rise to the scattering
equations, (this last fact was originally proposed by
Siegel in \cite{Siegel:2015axg})
\begin{equation}
  \label{eq:delta-taui}
  \int d \tau_i e^ {E_i \tau_i} \sim \delta(E_i)\,.
\end{equation}
with $E_i$ the term in the parenthesis in eq.~(\ref{eq:null-KN}).
Note that due to global GCA invariance, there are only $n-3$
independent GCA cross ratios and hence $n-3$ scattering equations.  In
our present case, with the conformal mapping
$\sigma\to \exp(i \sigma)$ used here, the scattering equations
appear first as
  \begin{equation}
    \label{eq:sc-eqn-cot}
    \forall i=1, \ldots, n-3\,, \quad E_i =    \sum_j  
    k_i \cdot k_j \left(\frac{1}{z_{ij}} +\frac{z_j}{z_i z_{ij}}\right)=0
  \end{equation}
  Using the partial fraction identity
  $\frac{z_j}{z_i
    z_{ij}}=\frac{1}{z_{ij}}-\frac{1}{z_i}$ and momentum conservation
  they reduce to the CHY scattering equations.
The extra factor of $z_i$ with $\tau_i$ in the exponential finally
ensures that the measure is invariant. When
$\tau_i\to \tilde \tau_i = z_i \tau_i$ and
$\sigma_i \to z_i=\exp(i\sigma_i)$:
\begin{equation}
  \label{eq:sigma-to-z}
  d \sigma_i d \tau_i \to d z_i d\tilde{\tau_i} 
\end{equation}
up to numerical factors of $2i\pi$.

\subsection{Partition function}
\label{sec:partition-function}
The operator formalism can also be used to give a tentative calculation of the partition function. Consider the trace
\begin{align}\label{partf}
 \mathcal{Z}(\rho,\phi)=\mathrm{Tr} (\exp(2\pi i\phi P - 2\pi\rho H)).
\end{align}
Here $P=L_0 - \frac{c}{24}$ is the generator of translations in space
along $\phi$, and $H=M_0$ is the Hamiltonian generating time evolution
along $\rho$. Here we have Wick rotated to Euclidean signature, hence
the absence of a factor of $i$ in front of the Hamiltonian.  A generic
state in the Hilbert space is given by polynomials of the negative
modes $y_n$ and $p_n$
\begin{align}
 |\phi_I\rangle = x^{\mu_1}_{a_1}\cdots x^{\mu_n}_{a_n} p^{\nu_1}_{b_1}\cdots p^{\nu_m}_{b_j}|k\rangle
\end{align}
where $|k\rangle=\exp(x_0\cdot k)|0\rangle$ is the vacuum with momenta $k$ and $I$ is a multi-index.

Acting with these translation operators on a generic state and tracing over gives
\begin{align}
 \mathcal{Z}(\rho,\phi)=\int \frac{dk}{(2\pi)}e^{-\pi\rho k^2}e^{2i\pi\phi c/24}\prod_{a=1}^{\infty}\prod_{b=1}^{\infty}\sum_{N_a=0}^{\infty}\sum_{N_b=0}^{\infty}e^{2\pi i aN_a\phi}e^{2\pi i bN_b\phi}
\end{align}
Performing the Gaussian integral and the sum we arrive at
\begin{align}
 \mathcal{Z}_1(\rho,\phi)=(4\pi^2\rho)^{-1/2}q^{c/24}\prod_{b=1}^{\infty}(1-q^b)^{-2}
\end{align}
where $q=e^{2\pi i\phi}$ is in principle a complex number of unit modulus. In the above we neglected the spacetime indices of the oscillators\footnote{We also threw out a dimension dependent overall constant which is basically the volume of a $D-1$ sphere.}, so in $D$ dimensions the partition function is 
\begin{align}\label{final_pf}
 \mathcal{Z}(\rho,\phi)=(4\pi^2\rho)^{-D/2}q^{c/24}\left(\prod_{b=1}^{\infty}(1-q^b)^{-2}\right)^D
\end{align}
Note how similar it is to the partition function of a (non-chiral) single boson
\begin{align}
 \mathcal{Z}_X=(4\pi^2\tau_2)^{-1/2}\left|q^{1/24}\prod_{n=1}^\infty(1-q^n)^{-1}\right|^2
\end{align}
but in this case $q$ is the modular parameter of the torus, not a unit
norm complex number as in the null string.

Comparing to the partition function of the ambitwistor string found in
\cite{Adamo:2013tsa} we see that there is an extra modulus, $\rho$, in
the null string. Furthermore, the modulus $q$ is the modular parameter
of the torus in \cite{Adamo:2013tsa} while in the null string it is a
complex number of unit norm. The ambitwistor string also has an
explicit integration over the zero mode of $P$, leading to a
loop-momentum integration. In the case of the null string the
loop-momentum integral is exchanged for an integral over the extra
modulus. We expect that it is this modulus $\rho$ which controls the
UV behaviour of the theory. From the previous sections we know that
the moduli space of the complexified null string is the cotangent to
the moduli space of Riemann surfaces. So it is natural to conjecture
that the moduli space of the real null string is a some real cycle in
this space. In fact, recent work in one-loop amplitudes in the null
string \cite{Yu:2017bpw} seems to support this hypothesis. The
partition function computed in this paper by different methods seems
to be the same as ours with a specific choice of contour.\footnote{The
  part that was computed there was the matter part; it matches our
  expression, up to numerical factors. The integration contour was,
  there also, conjectured.}

\subsection{Comment on modular invariance}
\label{sec:comm-modul-invar}
After complexifying we can imagine that the null string is a Galilean
conformal field theory obtained by contracting some CFT. Then the
parameters $(\rho,\phi)$ should inherit modular transformations from
the parent theory, see \cite{Bagchi:2013qva,Bagchi:2016geg,Bagchi:2017cpu}. With respect to the
parent CFT Virasoro, the GCA zero mode operators are
\begin{align}
 &L_0= \mathcal{L}_0 - \bar{\mathcal{L}}_0\nonumber\\
 &M_0=\epsilon(\mathcal{L}_0 + \bar{\mathcal{L}})\nonumber
\end{align}
Here, $\epsilon$ is a parameter that we will take to zero to perform
the algebra contraction. Call $\zeta,\bar\zeta$ the parameters
associated\footnote{In the sense of defining the partition function as
  above.} with $\mathcal{L}_0$ and $\bar{\mathcal{L}}_0$, respectively,
then the GCA parameters are $2\phi=\zeta + \bar\zeta$ and
$2\rho=\zeta - \bar\zeta$, associated to $L_0$ and $M_0$,
respectively. The parameter $\zeta$ and its complex conjugate are the
modular parameters of the torus carrying an action of the modular
group $SL(2,mathbb{Z})$
\begin{align}
 \zeta\rightarrow \frac{a\zeta+b}{c\zeta+d},\quad a,b,c,d\in\mathbb{Z},\,ab-dc=1
\end{align}
When taking the limit, $\rho$ scales as $\epsilon$ since it is associated with $M_0$. Making this explicit in the above and expanding to first order in $\epsilon$ gives
\begin{align}
 \phi+\epsilon\rho\rightarrow\frac{a(\phi+\epsilon\rho)+b}{c(\phi+\epsilon\rho)+d}
  = \frac{a\phi+b}{c\phi+d} +\epsilon \frac{\rho}{(c\phi+d)^2}
\label{eq:modular-squeeze}
\end{align}
The claim is then that the modular transformations for the null string are generated by
\begin{align}
 (\phi,\rho)&\rightarrow (\phi+1,\rho)\\
 (\phi,\rho)&\rightarrow \left(\frac{-1}{\phi},\frac{\rho}{\phi^2}\right)
\end{align}
With these transformations in hand we can examine how the partition function behaves under them. Rewriting it in terms of the eta-function $\eta(\tau)=q^{1/24}\prod_{n=1}^\infty(1-q^n)$ gives
\begin{align}
 Z_{XP}=(4\pi^2\rho)^{-D/2}(\eta(\phi))^{-2D}.
\end{align}
Under modular transformations the eta-function behaves as
\begin{align}
 &\eta(\tau+1)=\exp(i\pi/12)\eta(\tau)\nonumber\\
 &\eta(-1/\tau)=(-i\tau)^{1/2}\eta(\tau)\nonumber
\end{align}
It's clear that under these transformations $Z_{XP}$ picks up a phase. But all is not lost yet, so far we haven't included the ghost sector. Naively the partition function for the ghosts is just $\eta^4$. This is even worse since it picks up factors of $\phi$ under modular transformations. But the ghost partition function should not be taken into account without the anti-ghost insertions which builds the measure in the moduli space. Instead of deriving this measure we will assume modular invariance and show that it uniquely fixes the ghost partition function and the measure on the moduli space. The claim is that the ghost partition function is $\rho\eta(\phi)^4$ since this picks up a phase independent of $\phi$ under modular transformations. Combining these partition functions gives
\begin{align}
 Z_{XP}Z_g=(4\pi^2\rho)^{-D/2}(\eta(\phi))^{-2D}\rho(\eta(\phi))^4.
\end{align}
Its easy to see that the relative phases cancel when
$D=26$\footnote{That is sufficient, not necessary.}. There is also a
unique modular invariant measure in the space of $(\phi,\rho)$ which
combines with the partition function to give
\begin{align}
 \int \frac{d\phi d\rho}{(\rho)^2}(4\pi^2\rho)^{-13}(\eta(\phi))^{-52}\rho(\eta(\phi))^4
\end{align}
Note that in the above formula there was no need of assigning a
modular transformation to the field $P$ to get a modular invariant
function like in \cite{Adamo:2013tsa}. As expected, its role has been
taken over by the factor $(\rho)^{-13}$. An integrand that goes with
it also will not depend on the zero mode of $P$, but will depend on a new
modulus. Like the tree-level amplitude we expect this dependence to be
exponential which might allow for new loop-level scattering equations
without an explicit loop momentum.

As we mentioned, a one-loop amplitude in the bosonic null string has been
proposed in \cite{Yu:2017bpw}. Given a particular choice of contour
the authors recovered scalar boxes in Schwinger parametrization. It
would be very interesting to compute the one-loop amplitude using the
above operator formalism and compare with their results.

\section{Discussion}
\label{sec:discussion}

\paragraph{Summary}
In this paper, we pushed the study of the null string into three
different but related directions. First we complexified the worldsheet
and target space where we noticed an emergent symmetry which does not
preserve the original real contour. This symmetry is on-shell gauge
equivalent to holomorphic diffeomorphisms and corresponds to
translations along null geodesics which is the same as one of the gauge
symmetries of the ambitwistor string. In the same section we also
studied the moduli space of the null string and concluded it is the
same as the ambitwistor string when viewed through the lens of this
emergent symmetry.

Next we studied the role of the Galilean conformal algebra in the
structure of the null string. We showed how the constraints of the
null string restrict the state space to be the same as a chiral
CFT. This motivates why one can use the usual state-operator
correspondence in these models. Then we showed how the residual
symmetry acts locally and globally, and how to gauge fix it gives rise
to a Jacobian which matches with the ambitwistor string ghosts
correlator. In doing this we showed how the chiral gauge-fixing of the
ambitwistor string translates into the gauge fixing of the nulls
string and vice versa.

Lastly we looked at tree-level amplitudes using an operator
formalism. There we showed explicitly at four points how the extra
moduli of the null string can be used to obtain the tree-level
amplitudes in the CHY form, that is, localized to the scattering
equations. Next we calculated the cylinder propagator and gave an
n-point argument for how the scattering equations appear at
tree-level. We closed the section by calculating the partition
function from operator methods, pointing out its differences and
similarities with other ambitwistor partition functions in the
literature and showed that our partition function is invariant under a
conjectured action of modular transformations in the moduli space of
the null string.

\paragraph{Perspectives}
Going forward, there are many directions of research which this work
opens.

First, a full treatment of the path integral in the real setting, if
it makes sense, would be illuminating and might follow the lines
advocated in \cite{Sundborg:1994py}. The idea would be then to
determine the complex integration cycle (that are known as Lefchetz
thimbles \cite{Witten:2012bh,Ohmori:2015sha,Mizera:2017cqs}) by
computing the intersection between the real and complex case.

It would also be very interesting to understand the details of the
procedure sketched in section~\ref{sec:cylind-prop-n}. In particular,
it seems that there could be a choice in the order of integration,
$\tau$ or $z$ first. Even at tree-level doing so is difficult but
could lead to a new representation of the CHY formulae. At loop-level,
an interesting possibility arises, the loop momenta would naturally
arise within the scattering equations instead of being an explicit variable of integration. If it is possible to do the $z$
integral first, then the $\tau$ integral seems to reduce to a
Schwinger proper-time parametrization. Evidence for this was proposed
in \cite{Yu:2017bpw}.
However, we already mentioned that a lot of subtleties are present at
loop-level, and it is not at all obvious that such a thing is
possible. For this reason it will be necessary to understand further
the moduli space of the null string at loop-level.

Recently another proposal for a gauge-unfixed
version of the ambitwistor models in a first order
setting was put forward in~\cite{Arvanitakis:2017tla}. It argued that the resulting
models are essentially topological, and the BRST localization \cite{Ohmori:2015sha} of the
ambitwistor string on the scattering equations is essentially a kind of topological localization. It would be interesting to connect the two
approaches and put in perspective the earlier results of
\cite{Sundborg:1994py}.

Concerning supersymmetry, we mentioned that the analysis presented
here can be carried straightforwardly in the RNS model of
Mason-Skinner \cite{Mason:2013sva}, or in the pure spinor version of the formalism
\cite{Berkovits:2013xba,Adamo:2015hoa,Chandia:2016dwr}.

In our previous paper \cite{Casali:2016atr} we noticed that there are chiral models in which the tension is still present as a free parameter, these were later studied in \cite{Leite:2016fno,Huang:2016bdd,Siegel:2015axg,Azevedo:2017yjy}. It would be interesting to see if the methods developed in this paper can be applied to these models and how they relate to the null string and the usual string.

Finally, and in relation with the comment at the beginning of
section~\ref{sec:gca} on non-relativistic symmetries, it would be
interesting to see if there exist other type of string models which
could be quantized following the methods exposed in this paper. In
particular, as recalled in \cite{Duval:2014lpa}, Kar claimed in
\cite{Kar:1995br} that Schild's strings (by opposition to our LST
strings) enjoy a larger set of reparametrisations, spanning the full
Newman-Unti group. They are given by
$(\tau,\sigma)\to (f(\tau,\sigma),g(\sigma))$. It would be interesting
to study the quantization of these strings and see if they can be
related to LST strings.

\paragraph{Acknowledgements}
The authors would like to acknowledge Guillaume Bossard, Eric D'Hoker,
Lionel Mason for interesting discussions and comments.  All of us
would like to acknowledge hospitality from the Newton Institute and
the organizers of the program ``Gravity, twistors and amplitudes'' where
this work was initiated. EC and PT would like to acknowledge
hospitality form the Kavli Institute for Theoretical Physics and the
organizers of the program Scattering Amplitudes and Beyond during
which this work was partially finalized. This research was supported in part by
the National Science Foundation under Grant No. NSF PHY-1125915.  The
work of PT is supported by STFC grant ST/L000385/1.

\appendix

\section{Comments on worldline symmetries}
\label{sec:some-comm-worldl}

The equivalence between the antiholomorphic diffeomorphisms and
translations along null geodesics in the null string closely resembles
a similar phenomenon of the particle action. Here we review the
this equivalence in the worldline in order to illustrate what happens
in the null string.

The worldline action for a massless particle is written in second
order form as
\begin{align}\label{eq:part_second}
S=\int \sqrt{-g}g^{\tau\tau} (\partial_\tau x)^2. 
\end{align}
This action is invariant under diffeomorphism. Under $\tau\to\epsilon(\tau)$, the metric transforms as
\begin{equation}
  \label{eq:1}
  \delta g_{\tau\tau}=\epsilon \partial_\tau g_{\tau\tau} +
  2 g_{\tau\tau} \partial_\tau \epsilon,\quad \delta(\sqrt{g})=\partial_\tau(\epsilon \sqrt{g})
\end{equation}
We write the action in the first order formalism by introducing the canonical momenta $p$
$$ S_{(p,x)}=\int (p \partial_t x - \frac{e}2 p^2). $$
The corresponding equation of motion for $p$ is
$e^{-1}\partial_t x=p$.
Importantly, since the above action is equivalent to
\eqref{eq:part_second} it still is diffeomorphism invariant. Under
$\tau\to\tau+\epsilon(\tau)$ the fields transform as
\begin{equation}\label{eq:diffeo-wl}
\delta x=\epsilon \partial_t x,\quad
\delta p =\epsilon \partial_t p,\quad
\delta e= \partial_t(\epsilon e )
\end{equation}
where we identify $e= \sqrt{g_{\tau \tau}}$. However, it is also the
case that the gauge symmetries are generated by the constraints. In
this case the constraint $p^2$ generates the gauge symmetry of the system
\begin{equation}\label{eq:ll=wl}
\delta x=\alpha p,\quad
\delta p=0,\quad
\delta e = \partial_t \alpha\,.
\end{equation}
These two symmetries should be somehow equivalent, except that in the
Hamiltonian form we usually discard time parametrization, as these
are produced by changing the values of the Lagrange multipliers in the
extended Hamiltonian.

Henneaux and Teitelboim describe this phenomenon in their book
(\cite{Henneaux:1992ig}, chap. 3.1.5, ``Trivial gauge
transformations'') in some details. The important fact to notice here
is that the two symmetries just differ by a trivial ``equation of
motion symmetry''. In other words, an $\alpha$ transformation is
equal, on-shell, ($\partial_t p=0$) to a diffeomorphism, with
parameter $\epsilon=e \alpha$. The gauge transformation that is
obtained from the difference between these two is a trivial gauge
transformation. These trivial transformations that vanish on-shell can
always be written as~\cite[Thm 3.1]{Henneaux:1992ig}),
\begin{equation}
  \delta' y^i = \epsilon^{ij} \frac{\delta S}{\delta y^j}
\end{equation}
for canonical variables $y^i$ with action $S$ and, crucially, $\epsilon^{ij}$ some antisymmetric variable. In our case, (\ref{eq:diffeo-wl})-(\ref{eq:ll=wl}) gives 
\begin{equation}
  \label{eq:gauge-trivial}
  \begin{aligned}
    \delta' x =\epsilon (\partial_tx-e p) = \frac{\delta
      S_{(p,x)}}{\delta p}\\
    \delta' p = \epsilon \partial_t p = -\frac{\delta
      S_{(p,x)}}{\delta x}
  \end{aligned}
\end{equation}
These transformations form an ideal within the set of gauge transformation (their commutator with other always give another equation of motion symmetry). They should be disregarded, and a way to see this is that the associated charge is a function that vanishes identically.

Something very similar happens in the complexified null string. The
antiholomorphic or $\tau$ diffeomorphisms are equivalent on-shell to
the scaling symmetry present in the ambitwistor string generated by
the $P^2$ constraint.

It would be interesting to revisit this analysis using the light-front
formalism developed in \cite{Alexandrov:2014rta} to understand more
conceptually the constraint analysis presented here.

\section{Electrostatic equilibrium}

It was observed long ago that the scattering equations actually
describe an electrostatics equilibrium on the sphere
\cite{Gross:1987ar,Fairlie:1972zz}. We comment on this observation
from the point of view of the real null string.

Starting from the real LST action, the insertion of plane wave vertex
operators in the path integral, induces the addition of source terms
to the action, which play the role of boundary conditions in the path
integral:
\begin{equation}
  \label{eq:action-sources}
  \int d^2\sigma\left( V^\alpha V^\beta \partial_\alpha X \cdot \partial_\beta X + i \sum_{j=1}^n k_j\cdot X(\sigma,\tau) \delta^{(2)}(\sigma-\sigma_j,\tau-\tau_j)\right)\,.
\end{equation}
The corresponding $X$ equations of motion read
\begin{equation}
  \label{eq:eom-sources}
  \partial_\alpha (V^\alpha V^\beta \partial_\beta X^\mu) + i \sum_{j=1}^n k_j^\mu \delta^{(2)}(\sigma-\sigma_j,\tau-\tau_j)=0
\end{equation}
We want to interpret the following vector field as our electric field
(or rather a collection of electric fields, for $\mu=0,\dots,D-1$)
\begin{equation}
  \tilde E^\alpha_\mu=V^\alpha V^\beta \partial_\beta X_\mu
\end{equation}
This vector field has a density weight, which we can compensate by
introducing an auxiliary metric $g$ on the worldsheet, so we should
\begin{equation}
  \sqrt{-g}  E^\alpha_\mu=V^\alpha V^\beta \partial_\beta X_\mu
\end{equation}
so that $E$ is then a proper vector field. The equation of motion
\eqref{eq:eom-sources} then gives straight away Gauss's law in the
presence of sources. It would be interesting to work out the similar
configuration at loop level, pushing further the analysis of \cite{Casali:2014hfa}.

\newcommand*{\doi}[1]{\href{http://dx.doi.org/#1}{doi: #1}}
\bibliographystyle{JHEP} \bibliography{biblio}

\end{document}